\documentclass[12pt,english,twoside]{article}

\usepackage[T1]{fontenc}
\usepackage[utf8]{inputenc}
\usepackage[letterpaper, margin=2cm]{geometry}
\usepackage{color}
\usepackage{graphicx}
\usepackage{subfigure}
\graphicspath{{graphics_fold/}}
\usepackage{epstopdf} 
\usepackage{times}
\usepackage{etoolbox}
\usepackage{amsmath} 
\usepackage{amssymb}  
\usepackage{fancyhdr}
\usepackage{psfrag}
\usepackage{enumitem}
\usepackage{indentfirst}       
\usepackage{titlesec}          
\usepackage{hyperref}
\usepackage{babel}
\usepackage{listings}
\usepackage{booktabs}
\usepackage{tabularx}

\addto\captionsenglish{}

\usepackage[colorinlistoftodos]{todonotes} 

\setlist{nolistsep}

\makeatletter
\renewcommand*{\@biblabel}[1]{\hfill#1.}
\makeatother

\makeatletter
\patchcmd{\headrule}{\hrule}{\color{blue}\hrule}{}{}
\patchcmd{\footrule}{\hrule}{\color{blue}\hrule}{}{}
\makeatother

\fancypagestyle{firstpage}{
\fancyhead{}
\fancyfoot{}
\fancyfoot[LE,RO]{\small \thepage}

}

\makeatletter
\def\maketitle{
  \thispagestyle{firstpage}
\vspace*{-11mm}{\centering\includegraphics[width=\textwidth]{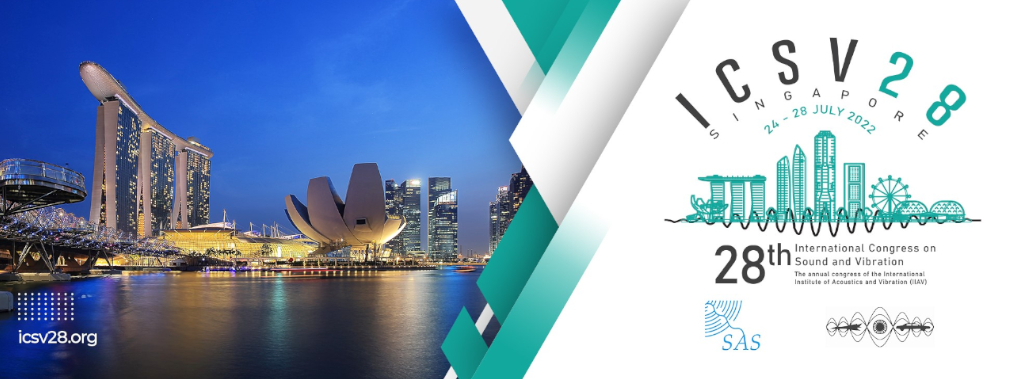}\\}
  {
   \vspace*{0mm}\fontsize{17}{20}\selectfont\sffamily{}  \noindent \MakeUppercase{\textbf{\@title}}

   \vspace*{3mm}\fontsize{14}{20}\selectfont\rmfamily{} \noindent \@author
  }
}
\makeatother

\titleformat{\section}
  {\fontsize{14}{14}\normalfont\sffamily\bfseries}
  {\thesection.}{1em}{}                

\titleformat{\subsection}
  {\normalfont\sffamily\bfseries}
  {\thesubsection}{1em}{}

\titleformat{\subsubsection}
  {\normalfont\sffamily\small}
  {\textit{\thesubsubsection}}{1em}\textit{{}}

\usepackage{amsfonts}

\usepackage[binary-units=true]{siunitx}
\usepackage{mathptmx}

\newcommand{\dBV}{$\si{\decibel} \slash \si{\volt}$}
\newcommand{\dBmW}{$\si{\decibel} \slash \si{\milli\watt}$}

\usepackage{placeins}

\title{PRELIMINARY ASSESSMENT OF A COST-EFFECTIVE\\ HEADPHONE CALIBRATION PROCEDURE FOR SOUNDSCAPE EVALUATIONS}

\author{%
    Bhan Lam,
    Kenneth Ooi,
    Karn N.\ Watcharasupat, 
    Zhen-Ting Ong,
    Yun-Ting Lau, \\
    Trevor~Wong, and
    Woon-Seng Gan\\
    {\small\textit{%
        School of Electrical and Electronic Engineering, Nanyang Technological University, Singapore\\
        50 Nanyang Avenue, Singapore 639798, Singapore\\
    e-mail: \{%
        bhanlam, wooi002, karn001, ztong, ylau01,  trevor.wong, ewsgan%
    \}@ntu.edu.sg
    }}
}

\begin{document}

\maketitle
\renewcommand{\abstractname}{\vspace{-\baselineskip}} 

\begin{abstract}	\noindent
The introduction of ISO 12913-2:2018 has provided a framework for standardized data collection and reporting procedures for soundscape practitioners. A strong emphasis was placed on the use of calibrated head and torso simulators (HATS) for binaural audio capture to obtain an accurate subjective impression and acoustic measure of the soundscape under evaluation. To auralise the binaural recordings as recorded or at set levels, the audio stimuli and the headphone setup are usually calibrated with a HATS. However, calibrated HATS are too financially prohibitive for most research teams, inevitably diminishing the availability of the soundscape standard. With the increasing availability of soundscape binaural recording datasets, and the importance of cross-cultural validation of the soundscape ISO standards, e.g.\ via the Soundscape Attributes Translation Project (SATP), it is imperative to assess the suitability of cost-effective headphone calibration methods to maximise availability without severely compromising on accuracy. Hence, this study objectively examines an open-circuit voltage (OCV) calibration method in comparison to a calibrated HATS on various soundcard and headphone combinations. Preliminary experiments found that calibration with the OCV method differed significantly from the reference binaural recordings in sound pressure levels, whereas negligible differences in levels were observed with the HATS calibration.  

\noindent Keywords: subjective listening; soundscape auralisation; playback calibration;
\end{abstract}

\quad\rule{425pt}{0.4pt}

\section{Introduction}

International standards with regards to the measurement of urban noise usually specify strict requirements for acoustic measurement requirements, for instance IEC 61672 standards for sound level meters and IEC 61904 standards for measurement microphones \cite{InternationalElectrotechnicalCommission2013IECSpecifications,InternationalElectrotechnicalCommission1996IECMicrophones}. This ensures that the physical quantities measured are traceable and reproducible. 
The ISO/TS 12913-2:2018 standard \cite{InternationalOrganizationforStandardization2018ISO/TSRequirements} on data collection and reporting guidelines for soundscape evaluations, specify minimum compliance with ITU-P P.58:2013 and ANSI/ASA S 3:36:2012 standards for binaural measurement systems \cite{AcousticalSocietyofAmericaASA2012ASA/ANSIMeasurements,InternationalTelecommunicationUnionRadiocommunicationSector2021ITU-TTelephonometry}. However, there is a lack of guidance on both the reproduction accuracy and calibration procedures concerning lab-based evaluation of soundscapes. Some research groups have thus resorted to calibration of soundtracks reproduced over headphones or loudspeakers with binaural systems that are compliant with ISO/TS 12913-2 \cite{Ooi2021AutomationHead} or with IEC 61672 Class 1 Sound Level Meters. 

Calibrated instruments, especially calibrated head and torso simulators (HATS), can be prohibitively expensive, which drastically reduces the accessibility of the ISO 12913 standards to less well-funded research groups. Notably, there is an increasing availability of open-access databases of audio-visual recordings that are compliant with ISO 12913, such as the ``Urban Soundscapes of the World'' project \cite{DeCoensel2017UrbanMind} and the ``International Soundscape Database'' \cite{Mitchell2021TheInformation}. Hence, there is a strong impetus to develop cost-effective calibration methods for lab-based reproduction of acoustic environments for soundscape assessments with minimal degradation of perceptual responses. 

For increased accessibility and inclusiveness, the ``Soundscape Attributes Translation Project'' (SATP) developed a cost-effective calibration methodology for headphones based on open-circuit voltage (OCV) measurements \cite{Aletta2020SoundscapeLanguages}. The goal of SATP is to generate validated translations of the perceived affective attributes in ISO 12913-2:2018 into as many spoken languages as possible \cite{Antunes2021ValidatedAssessment,Watcharasupat2022QuantitativeLanguage,Sudarsono2022TheStudy}. However, the proposed calibration method based on computing a theoretical root-mean-square (RMS) voltage from the manufacturer-provided sensitivity values could be undermined by a large variability in headphone quality, and sometimes unpredictable reproduction characteristics due to mismatched specifications between the headphones and audio output devices. This preliminary study sheds light on the variations in reproduced sound pressure levels (SPL) based on the OCV calibration method as compared to the HATS-based calibration method referencing the ISO 12913-2:2018 standard.

\section{Methodology}

Since the SATP initiative utilises a common set of audio stimuli to validate translations across all languages, this investigation on the calibration methods would be performed on this set of 27 stimuli. These 27 tracks were chosen to represent a diverse range of SPL levels, as well as spanning across the perceptual attribute space. The stimuli were recorded with a wearable binaural microphone (BHS~II, HEAD acoustics GmbH, Herzogenrath, Germany) and a data acqusition device (SQobold, HEAD acoustics GmbH, Herzogenrath, Germany) according to the protocol described in \cite{Mitchell2020TheInformation}.

\subsection{OCV Calibration Method}

Since the OCV calibration method\footnote{The OCV calibration method was originally developed for use by the SATP \cite{Aletta2020SoundscapeLanguages} working groups by Dr.\ Francesco Aletta, Dr.\ Tin Oberman, Andrew Mitchell, and Prof.\ Jian Kang, of the UCL Institute for Environmental Design and Engineering, The Bartlett Faculty of the Built Environment, University College London (UCL), London, United Kingdom.} is based on measurement of the root-mean-square (RMS) voltage at a single frequency, it inherently assumes that the entire system output chain responses linearly. In other words, the OCV method comes with a caveat that the audio output should be as linear as possible, e.g., via high-fidelity sound cards, and high-quality headphones with flat frequency responses. 

The headphone sensitivity in ``\dBV'', $S_\text{V}$, is a measure of how loud a headphone can produce a sound at a given RMS voltage, at a given frequency $f_0$. The OCV method leverages on the availability of this manufacturer-provided specification to compute the RMS voltage output that should be expected for any desired SPL. With the same ``reference'' track, this essentially calibrates the playback system (i.e., soundcard amplification, host PC volume output) to the headphones of the sensitivity $S_\text{V}$ used. It is also worthy to note that some manufacturers provide the sensitivity in terms of ``\dBmW'', which can be easily converted to \dBV using
\begin{equation} \label{eq:1}
   S_\text{V} = S_\text{mW} - 10\log_{10}\left(\frac{Z}{1000}\right),
\end{equation}
where $S_\text{mW}$ is the headphone sensitivity in \dBmW, and $Z$ is the impedance, in $\Omega$, of the headphones. It is worthy to note that $S_\text{mW}$ is in fact defined by
\begin{equation} \label{eq:2}
   S_\text{mW} = 10\log_{10}\left(\frac{\left(p\slash p_0\right)^2}{10^3 \cdot P}\right),
\end{equation}
where $p$ is the sound pressure in \si{\pascal}, $p_0=\SI{20}{\micro\pascal}$ is the reference sound pressure, and $P$ is the power in \si{\watt}. Hence, it implies that the headphones with $S_\text{mW}={96}{\text{\dBmW}}$ will produce \SI{96}{dBSPL} when $P=\SI{1}{\milli\watt}$, and the headphones would produce $96+10\log_{10}(2)\approx\SI{99}{dBSPL}$ at $P=\SI{2}{\milli\watt}$.

Using a reference track with a known SPL level $S_0$, in \si{dBSPL}, the required voltage can be calculated using the headphone sensitivity, $S_V$. This is done using a sine tone signal of frequency $f_0$, in \si{\hertz}, recorded at SPL of $S_0$.
To achieve the same SPL of $S_0$ with the headphone, the voltage $V$, in \si{\volt}, that has to be applied to the headphone jack can be calculated by
\begin{equation} \label{eq:3}
   20 \log_{10} V = S_0 - S_V.
\end{equation}

\begin{figure}[b]
    \centering
    \includegraphics[width=0.7\textwidth]{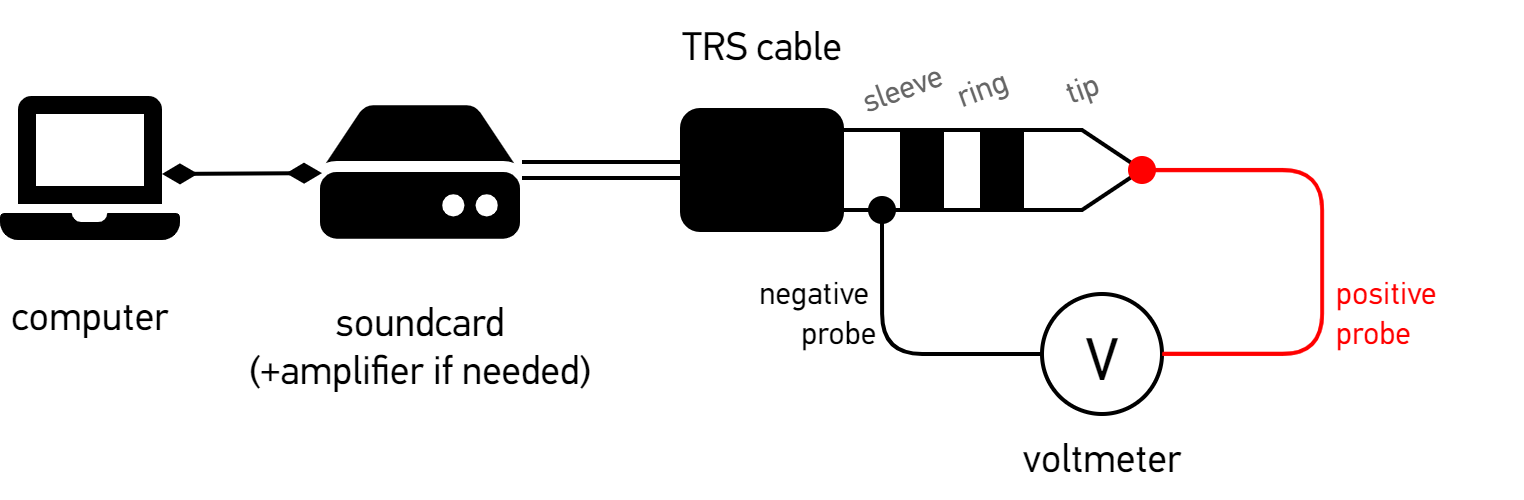}
    \caption{Illustration of the OCV method}
    \label{fig:ocv}
\end{figure}

For this work, a reference track of $S_0=\SI{94}{\decibel}$, $f_0=\SI{1}{\kilo\hertz}$ and a headphone (DT 990 Pro, Beyerdynamic GmbH \& Co.\ KG, Germany) with a sensitivity of $S_V=99.14$ was used. The required voltage computed with \eqref{eq:3} yields $V=\SI{0.553}{\volt}$, which should be the desired value measured at the output of the soundcard or headphone amplifier or whichever device the headphones will immediately receive the audio signal from. A voltmeter is used to measure the output of the aforementioned device when the reference track is playing, as shown in Figure \ref{fig:ocv}. Either the amplification on the soundcard or the volume of the host PC should be adjusted such that the voltmeter registers the desired levels, i.e. $V=\SI{0.553}{\volt}$. 

For headphones without detachable cables, a separate TRS cable with similar length should be used for the calibration process shown in Figure \ref{fig:ocv}. When the target voltage is observed and validated with multiple runs, the host PC volume levels and amplification levels of the sound car or headphone amplifier are considered to be calibrated and should be noted and fixed.

\subsection{HATS Calibration Method}

The HATS-based calibration method utilizes a calibrated HATS that complies minimally with relevant sections of the ITU-T P.58:2013 and ANSI/ASA S 3.36:2012 standards as defined in Annex D.6 of ISO 12913-2 \cite{InternationalOrganizationforStandardization2018ISO/TSRequirements}. This work employs an automated calibration setup described in \cite{Ooi2021AutomationHead} for the calibration of the audio tracks with a compliant HATS (45BB-5, G.R.A.S.\ Sound \& Vibration A/S, Holte, Denmark). The same pair of circumaural reference monitor headphones as that in the OCV method with a well-known neutral frequency response and high quality audio reproduction was used. The high-impedance headphones were driven by a high-fidelity soundcard (UltraLite~AVB, MOTU Inc, Cambridge, MA, USA). The calibration was performed in a custom soundproof chamber and controlled by National Instruments hardware and software as described in \cite{Ooi2021AutomationHead}. A summary of the hardware configuration for the HATS calibration is summarised in Table \ref{tab:hardware}.

\newcolumntype{Z}{>{\raggedright\arraybackslash}p{3.5cm}}
\newcolumntype{H}{>{\raggedright\arraybackslash}p{2cm}}
\newcolumntype{I}{>{\raggedright\arraybackslash}p{2.7cm}}
\newcolumntype{J}{>{\raggedright\arraybackslash}p{4.25cm}}
\newcolumntype{Y}{>{\raggedright\arraybackslash}p{4.25cm}}
\begin{table*}[tb]
\caption{Hardware specifications for calibration, playback, and recording of binaural audio tracks for soundscape evaluations}
\label{tab:hardware}
\setlength{\tabcolsep}{3pt}
\small
\begin{tabularx}{\textwidth}{HZJYI}
\toprule
Type                              
& Recommendations for calibration/playback                             & OCV                                                                  & HATS                                                                 & In situ
\\ \midrule

Headphones 
& Circumaural reference monitor headphones \cite{InternationalTelecommunicationUnionRadiocommunicationSector2015ITU-RSystems}                             
& Beyerdynamic DT 990 Pro & Beyerdynamic DT 990 Pro                     & NA                     
\\ \midrule
Soundcard 
& High-fidelity soundcard                                              
& MOTU Ultralite AVB & MOTU Ultralite AVB                         & NA                     
\\ \midrule
Head and torso simulator        
& Compliant with:
    \begin{itemize}[leftmargin=*]
      \item ITU-T P.58:2013, Section 5.2
      \item ANSI/ASA S3.36:2012, Table 1
      \end{itemize}
& NA
&GRAS 45BB-5 KEMAR Head and Torso
    \begin{itemize}[leftmargin=*]
      \item ANSI: S3.36, S3.25
      \item IEC: 60318-4
      \item ITU-T Rec. P.57 Type~3.3 based on ITU-T Rec. P.58
      \end{itemize}
& HEAD Acoustics BHS II  
\\ \midrule
Analog-to-digital converter (ADC) 
& Any ADC compliant with the head and torso simulator
    \begin{itemize}[leftmargin=*]
      \item Sampling rate: \SI{44.1}{\kilo\hertz} minimum
      \item Resolution: \SI{24}{bits} minimum
      \end{itemize}
& NA
& National Instruments
    \begin{itemize}[leftmargin=*]
          \item NI 9171
          \item NI 9234
          \end{itemize}
& HEAD Acoustics SQobold 
\\ \midrule
Acoustic environment              
& Compliant with ITU-R BS.1116–3, 8.2.1 \cite{InternationalTelecommunicationUnionRadiocommunicationSector2015ITU-RSystems}
& See \cite{Ooi2021AutomationHead} & See \cite{Ooi2021AutomationHead}     
& In situ                
\\
\bottomrule
\end{tabularx}
\end{table*}

During the calibration procedure, the headphones were placed over the ears of the HATS and the entire setup is sealed in the soundproof chamber (see \cite[Fig.\ 1]{Ooi2021AutomationHead}). The host PC digital volume levels and soundcard amplification were set to the same as that for the OCV calibration. For each of the 27 stimuli audio tracks, the calibration software searches for a digital amplification value such that the energetic average between the left and right channels are within a \SI{0.5}{dB} tolerance from the target SPL. The target SPL levels were the corresponding energetic averages of the left and right channels of each in-situ binaural recording. The calibration process was repeated for a total of three runs with repositioning of headphones between each run of the set of 27 tracks. 

\section{Results}
Before a comparison on reproduction accuracy can be made, the output levels of the OCV calibration method must first be measured. The same system setup and methodology used for the HATS calibration method was adopted for the measurement of the OCV-calibration headphone sound levels with a HATS. The automated calibration system \cite{Ooi2021AutomationHead} was configured to run without the search function and at a fixed gain of 1 to measure the SPL of the OCV calibration method as is. 

\begin{figure}[t]
    \centering
    \includegraphics[width=0.9\textwidth]{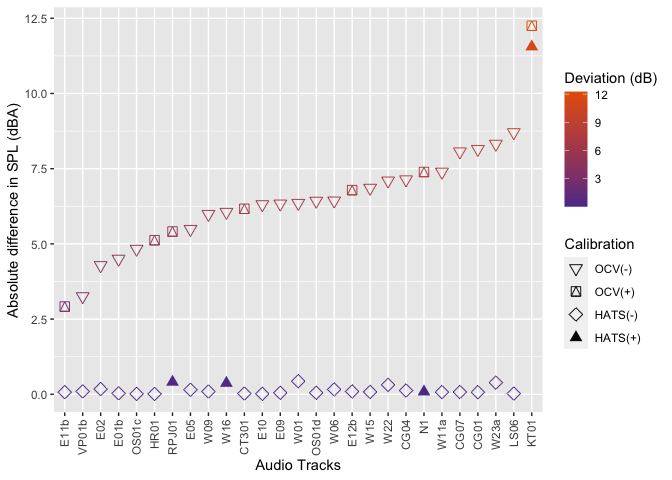}
    \caption{Absolute differences $\left\lvert\Delta_{\textsc{ocv}}\right\rvert$ and $\left\lvert\Delta_{\textsc{hats}}\right\rvert$ for each audio track, where suffix ``($-$)'' and ``($+$)'' represent negative and positive deviations with respect to the nominal values, respectively.}
    \label{fig:hats_calib}
\end{figure}

To illustrate the deviation in levels from the nominal values (i.e., of the in-situ binaural recording), the difference in A-weighted equivalent SPL across the duration of the entire sound track ($L_\text{A, eq}$) is computed for both the OCV and HATS calibration methods by 
\begin{equation} \label{eq:ocv}
    \Delta_{\textsc{ocv}} = L_\text{A, eq}^{(\textsc{ocv})} - L_\text{A, eq}^\text{(nom)},
\end{equation}
and
\begin{equation} \label{eq:hats}
    \Delta_{\textsc{hats}} = L_\text{A, eq}^{(\textsc{hats})} - L_\text{A, eq}^\text{(nom)},
\end{equation}
respectively. Both $\left\lvert\Delta_{\textsc{ocv}}\right\rvert$ and $\left\lvert\Delta_{\textsc{hats}}\right\rvert$ are plotted as a function of each audio stimuli in Figure \ref{fig:hats_calib}. 

Overall, $\left\lvert\Delta_{\textsc{ocv}}\right\rvert$ ranged from \SI{2.92}{\decibel} to \SI{12.25}{\decibel} with a mean and standard deviation of $\SI{6.45 \pm 1.85}{\decibel}$, as shown in Figure \ref{fig:hats_calib} and Table \ref{tab:SPL}. The OCV method resulted mostly in an under-calibration as indicated by the $\triangledown$ in Figure \ref{fig:hats_calib}. Except for track KT01, the $\left\lvert\Delta_{\textsc{hats}}\right\rvert$ was within the tolerance of $\pm\SI{0.5}{\decibel}$ as intended. The nominal SPL of KT01 ($L_\text{A, eq}^\text{(nom)}=\SI{40.19}{\decibel}$A) was below the noise floor of the measurement system ($\approx\SI{41}{\decibel}$A), which resulted in a failure of calibration. The mean and standard deviation of $\left\lvert\Delta_{\textsc{hats}}\right\rvert$ was $\SI{0.56 \pm 2.20}{\decibel}$ with KT01, and $\SI{0.140 \pm 0.132}{\decibel}$ without.

\begin{table}[bh!]
\centering
\caption{A-weighted equivalent sound pressure level values of the 27 tracks across in situ, OCV, and HATS calibration methods in dBA. The tracks are listed in the same order as Figure \ref{fig:hats_calib}}
\vspace{\baselineskip}
\setlength{\tabcolsep}{12pt}
\label{tab:SPL}
\begin{tabular}{%
l
*{3}{S[table-format=2.2]}
*{2}{S[table-format=+2.2]}
}
\toprule
{Track ID} & 
$L_\text{A, eq}^{(\text{nom})}$ &
$L_\text{A, eq}^{(\textsc{ocv})}$ &
$L_\text{A, eq}^{(\textsc{hats})}$ &
$\Delta_{\textsc{ocv}}$ &
$\Delta_{\textsc{hats}}$  \\ 
\midrule
E11b  & 85.94 & 88.86 & 85.87 & 2.92  & -0.07 \\
VP01b & 47.95 & 44.70 & 47.86 & -3.25 & -0.10 \\
E02   & 71.69 & 67.40 & 71.51 & -4.29 & -0.17 \\
E01b  & 66.74 & 62.24 & 66.71 & -4.50 & -0.03 \\
OS01c & 76.17 & 71.34 & 76.15 & -4.83 & -0.02 \\
HR01  & 73.42 & 78.54 & 73.40 & 5.12  & -0.01 \\
RPJ01 & 50.57 & 55.98 & 50.98 & 5.41  & 0.41  \\
E05   & 60.55 & 55.06 & 60.40 & -5.49 & -0.15 \\
W09   & 83.06 & 77.07 & 82.96 & -5.99 & -0.09 \\
W16   & 52.45 & 46.39 & 52.82 & -6.06 & 0.37  \\
CT301 & 84.91 & 91.08 & 84.89 & 6.17  & -0.03 \\
E10   & 75.38 & 69.07 & 75.36 & -6.31 & -0.02 \\
E09   & 67.93 & 61.60 & 67.88 & -6.34 & -0.05 \\
W01   & 73.09 & 66.74 & 72.66 & -6.36 & -0.44 \\
OS01d & 83.20 & 76.78 & 83.15 & -6.43 & -0.05 \\
W06   & 61.83 & 55.40 & 61.67 & -6.44 & -0.16 \\
E12b  & 76.24 & 83.03 & 76.15 & 6.79  & -0.09 \\
W15   & 64.92 & 58.06 & 64.84 & -6.86 & -0.08 \\
W22   & 59.82 & 52.72 & 59.51 & -7.10 & -0.31 \\
CG04  & 64.00 & 56.86 & 63.88 & -7.14 & -0.13 \\
N1    & 55.25 & 62.64 & 55.34 & 7.39  & 0.09  \\
W11a  & 66.02 & 58.62 & 65.94 & -7.39 & -0.07 \\
CG07  & 67.04 & 58.97 & 66.96 & -8.07 & -0.08 \\
CG01  & 73.34 & 65.19 & 73.27 & -8.15 & -0.07 \\
W23a  & 59.90 & 51.58 & 59.51 & -8.32 & -0.39 \\
LS06  & 72.30 & 63.59 & 72.27 & -8.71 & -0.03 \\
KT01  & 40.19 & 52.44 & 51.75 & 12.25 & 11.55 \\
\bottomrule
\end{tabular}
\end{table}

\section{Discussion and Conclusion}

The large deviation between the measured SPL of the OCV method and in-situ SPL may have potential implications on the subjective perception of the soundscapes. For instance, it was previously shown that an accurate reproduction of levels in a virtual reality display with binaural headphone playback was similar to the in situ experience in the perception of dominant sound sources and perceived affective quality attributes \cite{Hong2019QualityApplications}. Hence, the perceptual differences between OCV and HATS calibration methodologies should be investigated through subjective experiments and is currently ongoing.

The discrepancies in the OCV calibration method could be attributed to a high variability at \SI{1}{\kilo\hertz} even between calibrated HATS systems. Hence, HATS manufactures strongly recommend \SI{250}{\hertz} to be used as the calibration tone instead. Moreover, the true sensitivity values may vary between headphones of the same make and model, and also vary across frequencies within the same headphone. The sensitivity across $f_0$ is essentially the frequency response based on tonal excitation. If the headphones frequency responses are not flat, the calibration value will only be valid for the calibration frequency used, i.e., \SI{1}{\kilo\hertz}. 
 
Besides sensitivity, the impedance of the headphones are also known to independently influence the sound reproduction characteristics. For minimal loss of voltage or maximum transfer of voltage to the load (i.e., soundcard output to headphones), high impedance or voltage bridging is desired. Bridging is achieved by ensuring that the output impedance ($Z_\text{out}$) of the soundcard or amplifier is much lower than the impedance of the headphones ($Z_\text{in}$). Hence, reference monitor headphones with high impedance ($Z_\text{in}>\SI{50}{\ohm} \text{ to } \SI{100}{\ohm}$) should be paired to sources that are designed for high impedance loads. Likewise, low-impedance headphones (typically $Z_\text{in}\approx\SI{32}{\ohm}$) should be paired to sources with low output impedance (typically $Z_\text{out}<\SI{1}{\ohm}$). It should also be noted that some low $Z_\text{out}$ soundcards or amplifiers are also unable to drive low-impedance headphones due to limited available current. In general, high impedance headphones require lower current and vice versa. The reproduction characteristics can become unpredictable if the manufacturer does not provide output power specifications across a wide range of $Z_\text{in}$ or disclose current limiting mechanisms based on $Z_\text{in}$. Lastly, headphone impedance can vary significantly over frequencies. This directly affects voltage delivered to the headphone drivers at different frequencies. Although headphones with electrostatic drivers have an almost flat impedance response, they are prohibitively expensive and are not considered here. Therefore, a combination of the factors above could render the OCV method unreliable and as evidenced in the results, even for a headphone with flat impedance response and a properly matched soundcard. 

There is thus a need to investigate the perceptual differences between the OCV and HATS calibration method, as well as to determine if SPL deviations are consistent across a larger sample of headphone and soundcard combinations. 

\FloatBarrier

\section*{Acknowledgement}

The authors would like to thank Dr.\ Francesco Aletta, Dr.\ Tin Oberman, Andrew Mitchell, and Prof.\ Jian Kang, of the UCL Institute for Environmental Design and Engineering, The Bartlett Faculty of the Built Environment, University College London (UCL), London, United Kingdom, for development of the OCV calibration method.

This research is supported by the National Research Foundation, Singapore, and Ministry of National Development, Singapore under its Cities of Tomorrow R\&D Program (CoT Award: COT-V4–2020–1). Any opinions, findings and conclusions or recommendations expressed in this material are those of the authors and do not reflect the view of National Research Foundation, Singapore and Ministry of National Development, Singapore. 

\bibliographystyle{icsv_bib}
\bibliography{references,referencesBhan}

\end{document}